\documentstyle[procl,epsfig]{article}
\def\Journal#1#2#3#4{{#1} {\bf #2}, #3 (#4)}
\def\IJMP{{\em Int. J. Mod. Phys.} A}
\def\JETP{\em Sov. Phys. JETP}

\def\NPB{{\em Nucl. Phys.} B}
\def\PLB{{\em Phys. Lett.}  B}

\def\PRD{{\em Phys. Rev.} D}
\def\ZPC{{\em Z. Phys.} C}
\def\ibid{\em ibid.}
\def\citd#1#2{\cite{#1}$^{,\,}$\cite{#2}}
\def\citm#1#2{\cite{#1}$^{-\,}$\cite{#2}}
\def\half{{\textstyle {1\over2}}}
\def\third{{\textstyle {1\over3}}}

\def\as{\alpha_s}

\def\be{\begin{equation}}
\def\ee{\end{equation}}
\def\bea{\begin{eqnarray}}
\def\eea{\end{eqnarray}}
\def\cut{{\mbox{\scriptsize cut}}}
\def\jet{{\mbox{\scriptsize jet}}}
\begin{document}
\begin{flushright}
Cavendish--HEP--96/2\\
hep-ph/9607441
\end{flushright}
\title{DEEP INELASTIC SCATTERING --- THEORY AND PHENOMENOLOGY
\footnote{Talk at DIS96, International Workshop on Deep Inelastic
Scattering and Related Phenomena, Rome, April 1996.}}
\author{ B.R. WEBBER }
\address{Cavendish Laboratory, University of Cambridge,\\
Madingley Road, Cambridge CB3 0HE, England}

\maketitle\abstracts{
Recent developments in theory and phenomenology relevant to deep
inelastic lepton scattering are reviewed, concentrating on the
following topics: predicted behaviour of non-singlet and polarized
structure functions at small $x$; theoretical studies of saturation
and unitarity effects at small $x$ in quarkonium scattering;
renormalons and higher twist contributions;
next-to-leading-order calculations of jet cross sections;
forward jet production as a probe of small-$x$ dynamics.}

\section{Introduction}

The aim of this introductory review is to highlight some theoretical
and phenomenological work done in the past year or so which
may be relevant to deep inelastic scattering (DIS). The
selection of topics is of course subjective; it is intended
to cover a range of physics somewhat superficially, relying
on the talks to be presented in the working groups to fill
in the details.

In the important field of small-$x$ physics, I have chosen to
concentrate on rather non-standard topics: the small-$x$
behaviour of non-singlet and polarized structure functions,
and the small-$x$ dynamics of (quark)onium. This is because
there has been interesting new work in these areas, whereas
in the main arena of unpolarized singlet structure we are
experiencing something of a lull until more precise
calculations become available.

On the large topic of diffractive DIS I also have little to
say, although the study of multi-pomeron exchange in onium
scattering is relevant. The main developments in this area
are experimental at present and substantial
theoretical progress may be expected to come later.

I have included a rather speculative section on renormalons,
because this is a fashionable topic throughout QCD at present,
and it might just give some indication of where progress
could be made on the difficult problem of higher-twist
contributions.

Finally I discuss jet production, for which important new
calculations have been completed, although the situation
on their comparison with older calculations and with experiment
is still unclear.

\section{\boldmath Non-Singlet and Polarized Structure Functions at Small $x$}

There is a lot of interesting physics associated with the behaviour of
non-singlet and polarized structure functions, for example that
connected with sum rules. The Gottfried sum
\be
\int_0^1 \frac{dx}{x}(F_2^p - F_2^n)\simeq\third \left[
1+\int_0^1 dx (\bar u -\bar d)\right]
\ee
reveals that the $\bar u$ and $\bar d$ distributions are not the same,
while the Gross--Llewellyn-Smith and Bjorken sum rules
\bea\label{GLLSR}
\int_0^1 dx (F_3^\nu + F_3^{\bar\nu}) &=& 6\left[
1-\frac{\as}{\pi}-3.6\left(\frac{\as}{\pi}\right)^2
-19\left(\frac{\as}{\pi}\right)^3+\ldots\right]\;, \\
\label{BjSR}
\int_0^1 dx (g_1^p - g_1^n)&=& \frac 1 6 \left|\frac{g_A}{g_V}\right|
\left[1-\frac{\as}{\pi}-3.6\left(\frac{\as}{\pi}\right)^2
-20\left(\frac{\as}{\pi}\right)^3+\ldots\right]
\eea
provide precise determinations~\citd{Har95}{ElGaKaSa96}
of $\as$. The Ellis-Jaffe sum rule
for $g_1^p$ alone suggested the ``proton spin crisis".
All these sum rules require assumptions about how to extrapolate
non-singlet quantities towards $x=0$, so we need to understand
just what QCD predicts on this point.  This is an area where
there has been recent progress in theory and phenomenology.
\begin{figure}
\begin{center}
\epsfig{figure=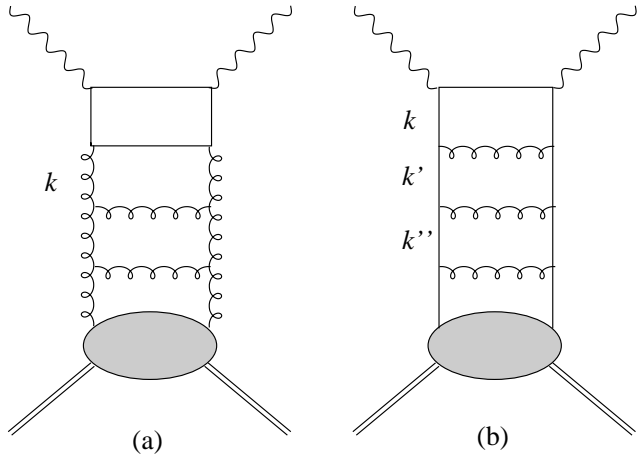,height=6.0cm}
\end{center}
\caption{(a) Gluon ladder and (b) quark ladder graphs, contributing to
singlet and non-singlet structure functions respectively.
\label{fig:ladders}}
\end{figure}

First let us recall what happens in the unpolarized, singlet case.
The behaviour is dominated by ladder graphs with longitudinally
polarized gluons on each side, Fig.~\ref{fig:ladders}(a),
which give a basic $x^{-1}$ dependence to the gluon distribution
$g(x)$ and hence to the structure functions. In DGLAP (leading-log $Q^2$)
evolution,\cite{DGLAP} this is modified by a relatively slowly-varying
asymptotic factor,
\be
g(x) \sim \frac 1 x \exp\left(\sqrt{4C_A\frac{\as}{\pi}
\ln\frac{Q^2}{Q_0^2}\ln\frac{1}{x}}\right)
\;\;\;\;\;\mbox{(DGLAP).}
\ee
The presence of two large logarithms for each power of $\as$ is
characteristic. Here one keeps only those terms in which one of
them is a $\log Q^2$. These come from the region of phase space
in which the transverse momenta $k_T$ of the exchanged gluons are
ordered along the ladder.

At small $x$ we need to keep instead the terms with the largest
number of factors of $\log x$. The $k_T$-ordering breaks down,
and there is the possibility of double-logarithmic
$(\as \ln^2 x)^n$ terms.  However, BFKL~\cite{BFKL} showed
that double logs of $x$ cancel in the fully inclusive,
unpolarized, singlet structure functions. Although
$k_T$ is not ordered, it diffuses without violent
fluctuations. This leads to the asymptotic behaviour
\be\label{eq:BFKL}
g(x) \sim \frac 1 x \exp\left(\chi_S C_A\frac{\as}{\pi}
\ln\frac{1}{x}\right)\sim x^{-1-\gamma_S}
\;\;\;\;\;\mbox{(BFKL).}
\ee
The leading singlet BFKL eigenvalue $\chi_S = 4\ln 2$ implies that
$\gamma_S\simeq 0.5$ for $\as\simeq 0.2$.

The observed behaviour is consistent with next-to-leading-order
(NLO) DGLAP evolution, and with the leading-order BFKL form (\ref{eq:BFKL})
for a reduced value of $\gamma_S$. It is expected that non-leading
corrections to the BFKL prediction will be important. They are
not yet known, but progress is being made on calculating them.\cite{FaLi}

\subsection{Non-Singlet Structure Functions at Small x}

For a non-singlet quantity $f(x)$, such as $(F^p_2-F^n_2)/x$,
we have quarks instead of gluons at the sides of the ladder,
as in  Fig.~\ref{fig:ladders}(b),
and leading log-$Q^2$ evolution now gives
\be
f(x) \sim \frac{1}{x^0} \exp\left(\sqrt{2C_F\frac{\as}{\pi}
\ln\frac{Q^2}{Q_0^2}\ln\frac{1}{x}}\right)
\;\;\;\;\;\mbox{(DGLAP).}
\ee
However, in this case the leading log-$x$ terms are
double-logarithmic.\citd{KiLi}{ErMaRy} 
Using a Sudakov representation of the exchanged momenta,
\be
k^\mu = xp^\mu + y\tilde q^\mu+k_T^\mu\;,
\ee
where $\tilde q=q+xp$ (so that $\tilde q^2=0$),
the double logarithms come from the region in which both momentum
fractions are strongly ordered, but in opposite directions:
\be
x\ll x'\ll x''\ll\cdots\;,\;\;\;
y\gg y'\gg y''\gg\cdots\;.
\ee
The dominant region of transverse momenta now becomes
\be
k_T^2/x >  {k'}_T^2/x' >  {k''}_T^2/x'' >\cdots\;.
\ee
Thus there can be huge fluctuations along the ladder, and
the breakdown of $k_T$ ordering is much more drastic than
in the singlet case. This leads to the Kirschner-Lipatov~\cite{KiLi}
(KL) type of small-$x$ behaviour:
\be
f(x) \sim \frac 1 {x^0} \exp\left(\sqrt{2C_F\frac{\as}{\pi}
\ln^2\frac{1}{x}}\right)\sim x^{-\gamma_{NS}^{(+)}}
\;\;\;\;\;\mbox{(KL)}
\ee
where $\gamma_{NS}^{(+)}=\sqrt{2C_F\as/\pi}\simeq 0.4$ for
$\as\simeq 0.2$.  Note that Regge behaviour~\cite{Hei} at small
$x$ would imply that
\be
f(x) \sim x^{-\alpha_\rho}\;\;\;\;\mbox{with}\;\alpha_\rho\simeq\half
\;\;\;\;\;\;\;\;\mbox{(R),}
\ee
which is similar.

\subsection{Polarized Structure Functions at Small x}

The structure function combination $\Delta g_1=g_1^p-g_1^n$
appearing in the Bjorken sum rule depends on the odd-signature
part of the forward virtual Compton amplitude (i.e.\ odd
under $s$-$u$ crossing). Therefore the expected Regge behaviour
is that associated with exchange of the odd-signature, isovector
$a_1$ trajectory,
\be
\Delta g_1(x)\sim
x^{-\alpha_{a_1}}\;\;\;\;\mbox{with}\;\alpha_{a_1}\simeq 0
\;\;\;\;\;\;\;\;\mbox{(R).}
\ee
Using the KL approach,\cite{KiLi} however, Bartels, Ermolaev and
Ryskin~\cite{BaErRy95} (BER) find that the corresponding power
$\gamma_{NS}^{(-)}$ is actually slightly {\em larger} than that
for even signature:
\be
\Delta g_1(x)\sim
x^{-\gamma_{NS}^{(-)}}>x^{-\gamma_{NS}^{(+)}}\sim x^{-0.4}
\;\;\;\;\;\mbox{(BER),}
\ee
and so the behaviour as $x\to 0$ is more singular.
\begin{figure}
\begin{center}
\epsfig{figure=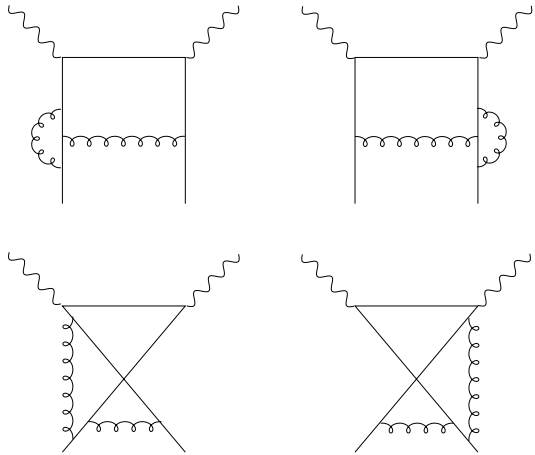,height=6.0cm}
\end{center}
\caption{Non-ladder contributions that cancel for even-signature
structure functions but not for odd signature.
\label{fig:nonladders}}
\end{figure}

The reason for the difference between even and odd signature is
that there are non-ladder contributions, like those in
Fig.~\ref{fig:nonladders}, which cancel in the former case but not in
the latter.  The even-signature cancellation follows from the fact
that the system exchanged in the $t$-channel is a colour singlet.
The surprising thing is that the non-ladder terms enhance the
odd-signature contribution rather than reducing it.
\begin{figure}
\begin{center}
\epsfig{figure=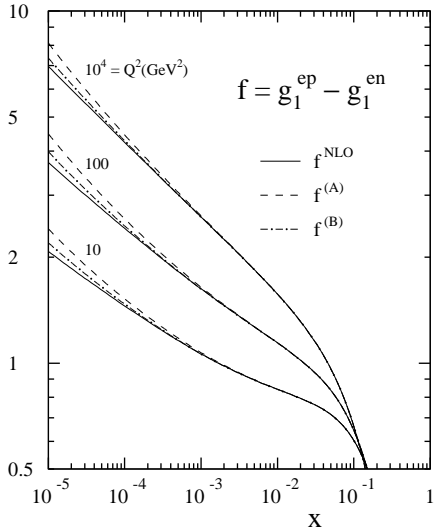,height=7.0cm}
\end{center}
\caption{Small-$x$ evolution of $g_1^p-g_1^n$. A and B denote different
prescriptions for fermion number conservation (see text).
\label{fig:g1diff}}
\end{figure}

An investigation of the phenomenological consequences for $\Delta g_1$
has been presented recently.\cite{BlVo}  Fig.~\ref{fig:g1diff}
shows the results of resumming the large logarithms of $x$ as
discussed above, compared with those of the standard next-to-leading
order DGLAP evolution. A difficulty with the resummation is that
the leading log-$x$ expressions do not conserve fermion number,
i.e.\ the integrals of the resummed splitting functions do not
vanish. One can fix this in various {\em ad hoc} ways which do
not affect the small-$x$ behaviour. The most extreme (curve A
in Fig.~\ref{fig:g1diff}) is to add a $\delta$-function
contribution at $x=1$. Curve B represents a less drastic
prescription.\cite{BlVo}  We see that the effect
in either case is not great at foreseeable values of
$x$ and $Q^2$.

The singlet contribution to $g_1$ at small $x$ has also been
considered.\cite{BaErRy96}  Here gluon ladders like
Fig.~\ref{fig:ladders}(a) can contribute,
but only one of the gluons on the sides can be longitudinal,
and so the basic power behaviour is $x^{-1+1} = x^0$, the same
as for quark ladders, Fig.~\ref{fig:ladders}(b).
Again, the odd signature of $g_1$
means that non-ladder contributions like those in
Fig.~\ref{fig:nonladders} do not cancel and in
fact enhance the structure function. The
asymptotic form is
\be
g_1(x) \sim \exp\left(\zeta_S\sqrt{C_A\frac{\as}{\pi}
\ln^2\frac{1}{x}}\right)\sim x^{-\gamma_{S}^{(-)}}
\;\;\;\;\;\mbox{(BER).}
\ee
The gluonic contribution alone would give $\zeta_S=4$ and
hence $\gamma_{S}^{(-)}=3\gamma_{NS}^{(-)}\simeq 1.2$.
Mixing with quark exchange reduces the leading
eigenvalue slightly:
\be
\zeta_S\simeq 3.5\;,\;\;\;\gamma_{S}^{(-)}\simeq 1.0\;,
\ee
implying that $g_1(x)\sim x^{-1}$. Further experimental data
on the small-$x$ behaviour of $g_1$ are needed to test this
surprising prediction.  The theoretical analysis~\cite{BaErRy96}
suggests other remarkable features, such as a possible change in
the sign of the polarized gluon distribution between large and small $x$.

\section{\boldmath Small-$x$ Saturation and Unitarity in
  Onium--Onium Scattering}

Amongst the important new phenomena expected in deep inelastic scattering
at small $x$ are parton saturation and unitarity corrections.\cite{GrLR83}
As the parton density increases, interactions between partons become more
significant and limit the rate of growth of the density (saturation).
In addition, multiple scattering effects become necessary in order
to prevent cross sections from exceeding unitarity limits.  These
phenomena are difficult to predict reliably in the case of $ep$
scattering because the proton is a large object whose structure
cannot be computed perturbatively. To study them within the domain
of perturbation theory we need to consider a smaller system such as
a heavy quark-antiquark bound state. For a sufficiently large quark
mass $m_Q$, the size of such a state is $R\sim 1/\as m_Q$.
Then the strong coupling at the relevant momentum scale
$q^2\sim 1/R^2$ will be small enough for a perturbative
expansion in $\as(q^2)$ to be applicable.

\begin{figure}
\begin{center}
\epsfig{figure=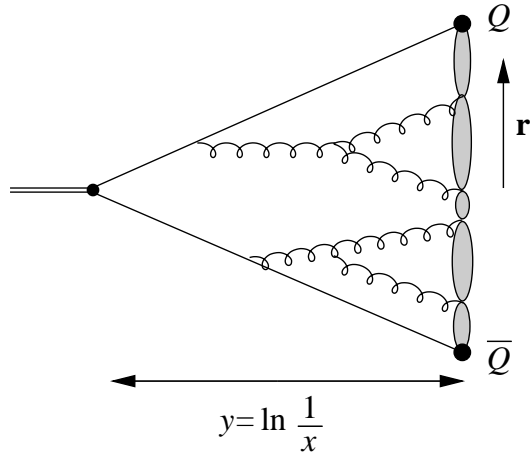,height=6.0cm}
\end{center}
\caption{Development of dipole chain in onium wave function.
\label{fig:chain}}
\end{figure}

The theory of low-$x$ phenomena in quarkonium (often simply called
onium) has been worked out in some detail over the last couple of
years using the colour dipole formulation~\citm{Muel94}{NaPeRo96}
of QCD at small $x$. The dipole formulation is based on the fact that,
to leading order in the number of colours, the light-cone
wave function of an onium state can be represented by a chain of colour
dipoles linking the heavy quark and antiquark (Fig.~\ref{fig:chain}).
As the wave function is probed at smaller $x$, i.e. higher `rapidity'
$y=\ln(1/x)$, more and more dipoles are likely to be resolved.
The evolution of the dipole configurations with increasing $y$
can be formulated as a branching process governed by
the BFKL evolution equation. It turns out to be most convenient
to describe the wave function in a (transverse position, rapidity)
({\bf r},$y$) representation. The mean density of dipoles of
size $c$ at a distance $r$ from a primary onium ($Q\bar Q$)
dipole of size $b$ is found to be
\be
\bar n(b,c,r,y)\sim \exp\left[(\alpha_P-1)y-\frac{\ln^2(16r^2/bc)}{ky}
\right]
\ee
where
\be
\alpha_P = 1+4\ln 2\,C_A\as/\pi
\ee
is the BFKL pomeron intercept and
\be
k = 14\zeta(3)\,C_A\as/\pi
\ee
is the corresponding diffusion constant. Thus as $y$ increases
the dipole density grows in the expected manner and dipoles diffuse
slowly (logarithmically in {\bf r}) from the onium source. The
growth of dipole density leads to saturation and unitarity
effects that can be investigated using the `theoretical
laboratory' of onium physics.\footnote{Applications of the
dipole formulation to DIS phenomenology are also being
investigated.\citm{NiZZ94a}{NaPeRo96}}

\subsection{Onium--Onium Scattering}

To study the effects of unitarity it is simplest to consider
onium-onium elastic scattering~\citm{Muel94}{MuSal96}
at very high energy $\sqrt s$.
The onia pass through each other with relative rapidity
$Y\sim \ln(s/m_Q^2)$ at some relative impact parameter {\bf r}.
In the centre-of-mass frame, the dominant elastic process
is two-gluon exchange between dipoles in the two onium wave
functions evolved to rapidity $y=Y/2$. This corresponds to
exchange of a single BFKL pomeron: the two evolved wave
functions provide the upper and lower halves of a gluon
ladder, which are joined in the middle by the two exchanged
gluons. The two-gluon exchange interaction is short-ranged
(it falls off like $1/r^4$ at large transverse distances),
and so only those dipole that overlap
contribute significantly to the scattering amplitude.
\begin{figure}
\begin{center}
\epsfig{figure=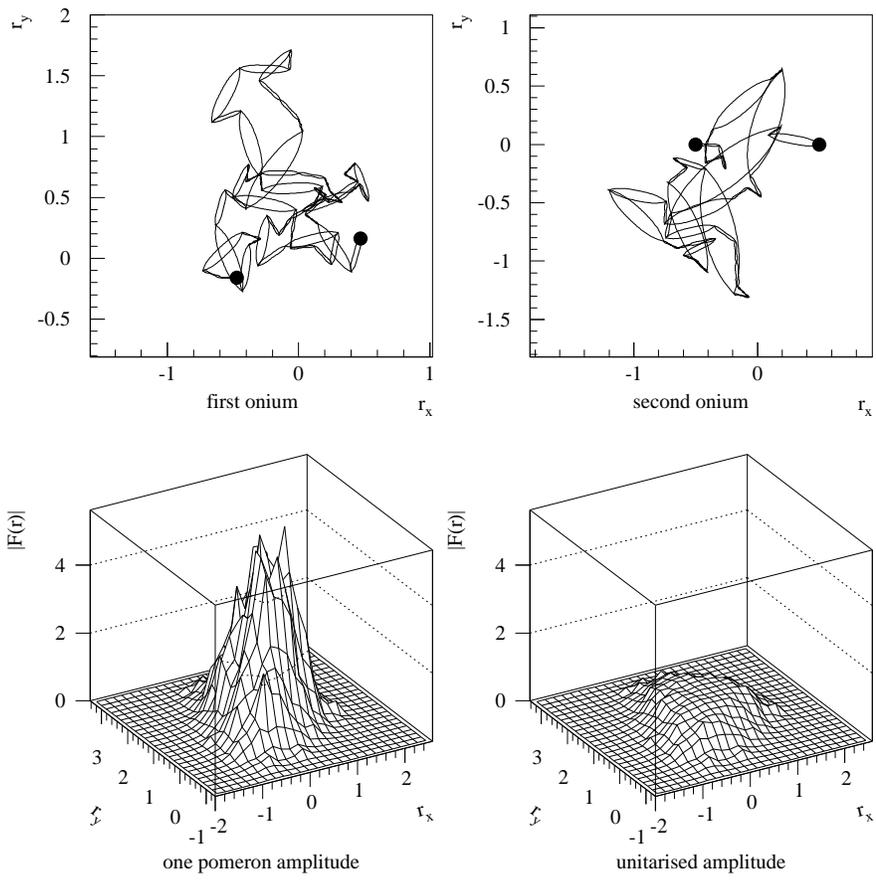,height=13.0cm}
\end{center}
\caption{Typical onium dipole configurations and their
one-pomeron and unitarized interaction amplitudes.
\label{fig:eg_event}}
\end{figure}

Fig.~\ref{fig:eg_event} illustrates the situation for the two typical
onium configurations shown at the top.\footnote{This figure and the
results that follow were obtained by Salam~\cite{Sala95} using his
Monte Carlo program OEDIPUS~\cite{SalaMC} (Onium Evolution, Dipole
Interaction, Perturbative Unitarization Software).}
The impact parameter profile
of the one-pomeron amplitude, shown lower left, is obtained by moving
one configuration over the other and computing the sum of the two-gluon
exchange amplitudes between all pairs of dipoles at each relative
position. Thus at impact parameters for which many dipoles overlap
the one-pomeron amplitude violates the unitarity limit, as shown by
the spikes exceeding unity.

\subsection{Unitarity and Saturation}

\begin{figure}
\begin{center}
\epsfig{figure=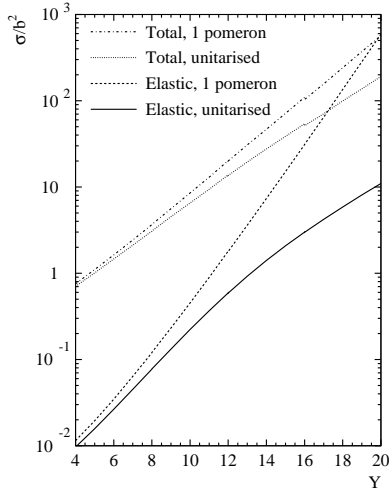,height=7.0cm}
\end{center}
\caption{Onium-onium total and elastic scattering cross sections.
\label{fig:tot_el_sigma}}
\end{figure}

Schematically, the elastic $S$-matrix element for single pomeron exchange
has the form
\be
S = 1-\sum_{\gamma,\gamma'} P_\gamma P_{\gamma'} f_{\gamma,\gamma'}
\ee
where $\gamma$ and $\gamma'$ represent dipole configurations of the
two onia with probabilities $P_\gamma$ and $P_{\gamma'}$ respectively,
and $if_{\gamma,\gamma'}$ is the one-pomeron amplitude for
those configurations. To satisfy unitarity we need to take into
account multiple pomeron exchange. For large dipole multiplicities
the $n$-pomeron amplitude exponentiates, i.e.
\be\label{Ssum}
S = 1-\sum_n\sum_{\gamma,\gamma'} P_\gamma P_{\gamma'}
f^{(n)}_{\gamma,\gamma'}
\ee
where
\be
f^{(n)}_{\gamma,\gamma'} \sim -\frac{(-f_{\gamma,\gamma'})^n}{n!}
\ee
and hence
\be\label{Sexp}
S \sim\sum_{\gamma,\gamma'} P_\gamma P_{\gamma'}\exp(-f_{\gamma,\gamma'})\;.
\ee
This ensures that the elastic amplitude never exceeds the unitarity
limit, as illustrated for the particular configurations in
Fig.~\ref{fig:eg_event} by the profile on the lower right.
The resulting changes in the total and elastic cross sections
as functions of rapidity are shown in Fig.~\ref{fig:tot_el_sigma}.

We see that the one-pomeron total and elastic cross sections have
the expected energy dependences, $s^{\alpha_P-1}$ and $s^{2(\alpha_P-1)}$
respectively. The unitarized total cross section has a somewhat
reduced effective power, while the elastic is more strongly affected.
This is because elastic scattering is more dominated by smaller
impact parameters, where large dipole densities are more probable.

When one performs the sum over multiple pomeron exchanges first,
and then the sum over dipole configurations, as in Eq.~(\ref{Sexp}),
it is obvious that a stable, unitary result is obtained. On the
other hand one may evaluate the terms in Eq.~(\ref{Ssum})
separately for each $n$, by summing over configurations first.
However, the multiple scattering series so obtained is
strongly divergent.\citd{Muel94}{Sala95} The reason is that
the configuration probability distributions $P_\gamma$ and
$P_{\gamma'}$ have long (exponential) tails at high dipole
multiplicities, which become dominant at high $n$ but are
strongly suppressed after summation over $n$.  This
suggests that for light hadrons also an expansion in
terms of the number of pomerons exchanged may not be
useful at very high energies.

So far, we have discussed everything in the onium-onium centre-of-mass
frame, where the two onia evolve through equal rapidity intervals,
$y=Y/2$.  We could equally well consider a `fixed-target' type of frame,
in which one onium evolves through $y\sim Y$ and the other remains close
to a pure $Q\bar Q$ configuration. One can show that the above treatment
is frame-independent at the level of one-pomeron (two-gluon) exchange,
but the results for multiple exchange are frame dependent. The reason
is that multiple scattering in the c.m.\ frame looks like parton-parton
interaction (saturation) in the wave function of the moving onium as
seen from the rest frame of the other. Thus unitarity and saturation
effects are intimately related, and one can learn about saturation
by enforcing the condition that the simultaneous treatment of the
two phenomena should be frame-independent.\cite{MuSal96}
The reader is referred to the talk by Salam~\cite{SalRom}
for further details. 


\section{Renormalons and Higher Twist}

An important thing to bear in mind when computing predictions of
perturbative QCD is that the perturbation series is not expected
to converge.  There may already be indications of this in some
quantities relevant to DIS: for example, the perturbative coefficients
in the Gross--Llewellyn-Smith and Bjorken sum rules, Eqs.~(\ref{GLLSR})
and (\ref{BjSR}), are growing rapidly.

\begin{figure}
\begin{center}
\epsfig{figure=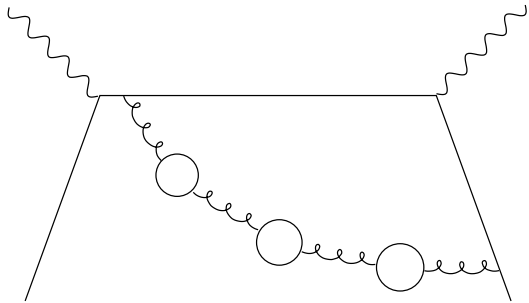,height=4.0cm}
\end{center}
\caption{A graph containing a ``renormalon chain''.
\label{fig:renormalon}}
\end{figure}

A known source of divergence of the perturbative expansion
is the set of so-called {\em renormalon} graphs,\cite{renormalons}
illustrated in Fig.~\ref{fig:renormalon}.
Summing chains of vacuum polarization bubbles on the gluon line leads
to a prediction of the form
\be\label{facdiv}
F = \sum_n c_n \as^n
\ee
where the coefficients $c_n$ are factorially divergent at high orders:
\be\label{cns}
c_n\sim n!\,(b/p)^n\;,
\ee
$p$ being a rational number that depends on the particular quantity $F$
being evaluated. Including $n_f$ flavours of fermionic bubbles like those
shown in Fig.~\ref{fig:renormalon}, one finds $b=-n_f/6\pi$.  To take
account of gluonic bubbles as well, the `naive non-Abelianization'
procedure~\citm{Ben94}{Bal95} is generally adopted: one evaluates the
graphs with quark bubbles and then simply makes the replacement
\be
n_f \to n_f-\frac{33}{2}\;,
\ee
i.e. one uses for $b$ the full 1-loop QCD $\beta$-function coefficient
$b=(33-2n_f)/12\pi$.

In fact the $\beta$-function in QCD (unlike QED) cannot be associated
with vacuum polarization graphs alone, and there is no set of graphs
that corresponds precisely to Eq.~(\ref{facdiv}) with
$b=(33-2n_f)/12\pi$. Nevertheless the naive non-Abelianization
procedure seems to work quite well in predicting the high-order
coefficients in perturbation theory. An impressive example
is the weak-coupling expansion of the average plaquette in quenched
($n_f=0$) lattice QCD.\cite{DiOnMa95}  For this quantity the
perturbative coefficients $c_n$ in Eq.~(\ref{facdiv}) up to $n=8$,
suitably re-interpreted to allow for the difference between the
lattice and continuum definitions of $\as$, have been
shown~\cite{DiOnMa95} to be dominated by a renormalon
contribution of the form (\ref{cns}) with $b=11/4\pi$ and $p=2$.

Although the series (\ref{facdiv}) is divergent, it is not useless:
it is expected to be an asymptotic expansion. By this we
mean that the series truncated at order $n=m$ is an approximation to
the true value of $F$, with an error bounded by the last term
retained, $c_m \as^m$. Now the coefficients $c_n$ satisfy
\be
\frac{c_n}{c_{n-1}}\sim n\frac{b}{p}
\ee
and thus the smallest term occurs at $n=m\sim p/b\as$.
Truncating the series at this point, we obtain the minimal
ambiguity in the prediction of $F$, which is
\be
\delta F \sim m!\, m^{-m}\sim e^{-m}\sim\exp\left(-\frac{p}{b\as}\right)\;.
\ee
Noting that
\be
\as\sim \frac{1}{b\ln (Q^2/\Lambda^2)}
\ee
where $Q^2$ is the relevant hard scattering scale, we see that
\be\label{dFpower}
\delta F \sim \left(\frac{\Lambda^2}{Q^2}\right)^p\;.
\ee
We may distinguish two cases:
\begin{enumerate}
\item $p<0$. In QCD, this is an {\em ultraviolet} renormalon.
The divergent series (\ref{facdiv}) has alternating signs.
It might appear from Eq.~(\ref{dFpower}) that this will lead
to an ambiguity in $F$ that grows with
increasing $Q^2$. In fact, however, the series can be summed
unambiguously in this case and there is no renormalon
ambiguity.

\item $p>0$, which is called an {\em infrared} renormalon. All the
terms of the divergent series have the same sign and it cannot be
summed unambiguously. The ambiguity (\ref{dFpower}) decreases
like $Q^{-2p}$ with increasing $Q^2$. 
\end{enumerate}

The above general discussion has some interesting possible applications
to deep inelastic scattering. In the case of ultraviolet renormalons,
various techniques can be used to sum the divergent terms and thus
improve the precision of the perturbative
prediction.\citm{LoMax95}{Bal95} Infrared
renormalons, on the other hand, imply that the perturbative
prediction must be supplemented by additional non-perturbative
information. In DIS, the power-suppressed nature of the ambiguity
suggests that the extra information required concerns {\em higher-twist}
contributions. In fact one can show~\cite{Muel93} that similar
renormalon graphs generate an ambiguity in the coefficient of
the leading $Q^{-2p}$ (twist $2p+2$) contribution, which cancels
the power-suppressed ambiguity in the twist-2 prediction.

\begin{figure}
\begin{center}
\epsfig{figure=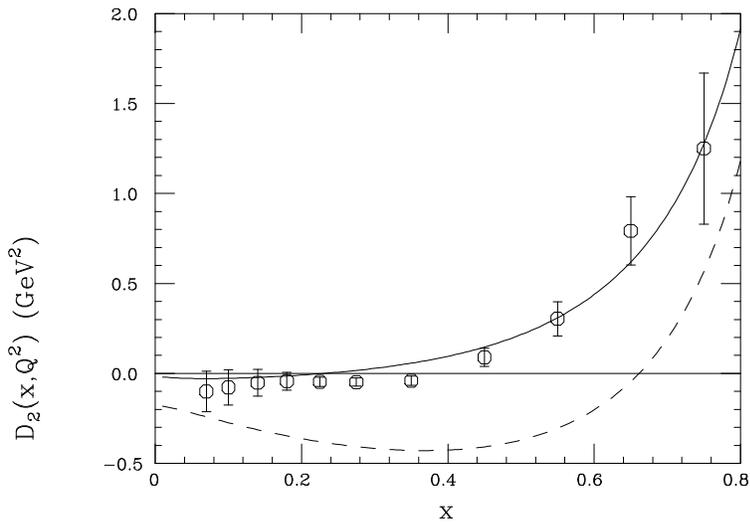,height=7.0cm}
\end{center}
\caption{Power corrections to $F_2$ (solid) and $xF_3$ (dashed). Points are
values deduced from BCDMS and SLAC data on $F_2$ (see text).
\label{fig:dispow}}
\end{figure}
Thus the presence of an infrared renormalon signals the need for
a higher-twist contribution and tells us what its power dependence
on $Q^2$ must be. This suggests that we could go further and
explore the possibility that the Bjorken-$x$ dependence of the
renormalon also indicates the $x$-dependence of the higher-twist
contribution. Phenomenologically, this appears to be the
case.\citm{Braun95}{DasWeb96}  Fig.~\ref{fig:dispow} shows,
for example, predictions~\citd{BPY95}{DasWeb96} based on this
idea for the $1/Q^2$-suppressed corrections to the structure
functions $F_2$ and $xF_3$, expressed in the form
\be\label{Ftwist}
F(x,Q^2) \simeq xq(x,Q^2)\left(1+\frac{D_2(x,Q^2)}{Q^2}\right)
\ee
where $q$ is the appropriate combination of quark distributions.
The form of the coefficient $D_2$ for $F_2$ is similar to that
deduced~\cite{VirMil92} from BCDMS~\cite{BCDMS} and SLAC~\cite{SLAC}
data, shown by the points.
Recent results~\cite{Sid96} on higher twist in $xF_3$ also look
similar in shape to the relevant curve (dashed).  It would clearly
be of interest to include the type of higher-twist contributions
suggested by this approach in global fits to deep inelastic
scattering data.

\section{Jet Production}

The important topic of jet production in DIS has also seen
recent theoretical and phenomenological progress. There
are new results from complete next-to-leading-order (NLO)
calculations of jet rates, which will be useful for
improving measurements of the strong coupling~\cite{exp_as}
and the gluon distribution.\cite{exp_glu}
In addition there are new calculations on testing BFKL dynamics
at small $x$ with forward jet production as a trigger.

\subsection{Jet Cross Sections in Next-to-Leading Order}

At present there are two programs in use to compute the two-jet
(plus beam remnant) cross section in NLO using a JADE~\cite{Jade}
type of jet definition: PROJET~\cite{Projet} and DISJET.\cite{Disjet}
There is now a new program, MEPJET,\cite{Mepjet} which can be used
to compute jet cross sections using any infrared-safe definition.
In addition there is the promise of a general-purpose program for the
NLO calculation of any infrared-safe quantity in DIS.\cite{CatSey96}

\begin{table}[h]
\begin{center}
\caption{Predicted two-jet inclusive cross sections in DIS at LO and NLO,
for four jet definition schemes and (at NLO) three 
different recombination schemes.
}\label{table1}
\vspace{3mm}
\begin{tabular}{|lcccc|}
\hline
& & & & \\
 Scheme
     &  \mbox{LO}
     &  \mbox{NLO} ($E$)
     &  \mbox{NLO} ($E0$)
     &  \mbox{NLO} ($P$)\\
\hline
$W$         & 1020 pb 
& 2082 pb & 1438 pb  & 1315 pb \\
\mbox{JADE} & 1020 pb 
& 1507 pb & 1387 pb  & 1265 pb \\
$k_T$       & 1067 pb 
& 1038 pb & 1014 pb  & \ 944 pb \\
\mbox{cone} & 1107~pb 
& 1203~pb & 1232 pb  & 1208 pb \\
\hline
\end{tabular}\end{center}
\end{table}
Results~\cite{MirZep96} on the inclusive two-jet cross section using
MEPJET with a variety of jet definitions, for a particular set of
acceptance cuts at HERA, are summarized in Table~\ref{table1}.
The jet-defining scheme that has been most widely used for
NLO calculations is the so-called $W$-scheme,\cite{Wscheme}
in which two particles or jets are clustered if their invariant
mass-squared $s_{ij}$ is less than $y_\cut W^2$, where
$y_\cut$ is the jet resolution ($y_\cut= 0.02$ here)
and $W$ is the hadronic c.m.\ energy.
Experimentally, the JADE clustering
variable $M^2_{ij}\equiv 2E_iE_j(1-\cos\theta_{ij})$ has been
used in place of $s_{ij}$. A suggested improvement is the
$k_\perp$-scheme,\cite{ktalg} in which
$k^2_{\perp,ij}\equiv 2\min\{E_i^2,E_j^2\}(1-\cos\theta_{ij})$,
evaluated in the Breit frame of reference, is compared with
a resolution scale (40 GeV$^2$ here). Alternatively,
one may use instead a cone type of jet definition, in which
clustering occurs if $(\eta_i-\eta_j)^2 +(\phi_i-\phi_j)^2 < 1$,
where $\eta$ and $\phi$ are the pseudorapidity and azimuth in
the HERA lab frame.  In all these schemes the way of defining the
combined momentum after clustering can also be varied, leading to
the so-called $E$, $E0$ and $P$ recombination schemes.\cite{BeKuSoSt}

It can be seen from Table~\ref{table1} that in the $W$ and JADE schemes
there are large values of the $K$-factor ($K \equiv$ [NLO-LO]/LO) and
strong recombination scheme dependences, which are reduced
in the $k_\perp$ and cone schemes.  One worrying point is that the
MEPJET results do not agree with those of PROJET and DISJET for the
$W$ or JADE schemes.\cite{MirZep96} This makes it very desirable
to have another independent calculation of these quantities, which is
said to be coming in the near future.\cite{CatSey96a}

\subsection{Forward Jet Production}

\begin{figure}
\begin{center}
\epsfig{figure=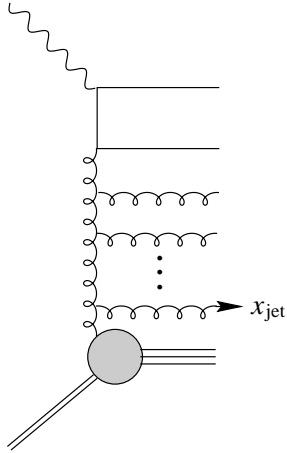,height=6.0cm}
\end{center}
\caption{Forward jet production as a probe of BFKL dynamics.
\label{fig:bfkl_jet}}
\end{figure}

The rate of production of fast ($x_\jet\gg x$) forward jets in
DIS at small $x$ has been proposed as a good probe of BFKL
dynamics.\citd{Muel90}{KwiMarSut}  The idea (Fig.~\ref{fig:bfkl_jet})
is that for jet transverse momenta $p^2_{T,\jet}\sim Q^2$ there is
little scope for ordinary DGLAP evolution, while the condition
$x_\jet\gg x$ means that BFKL evolution in log $x$ is enhanced.

In order to show that there is a BFKL enhancement
in forward jet production at small $x$, one should establish that
the usual NLO calculation with DGLAP evolution is not able to
explain the data. This is difficult because
NLO processes like boson-gluon fusion with an extra jet
are dominant in the selected kinematic region.\cite{MirZep96}
Thus NLO corrections are large and not well under control.

\begin{figure}
\begin{center}
\epsfig{figure=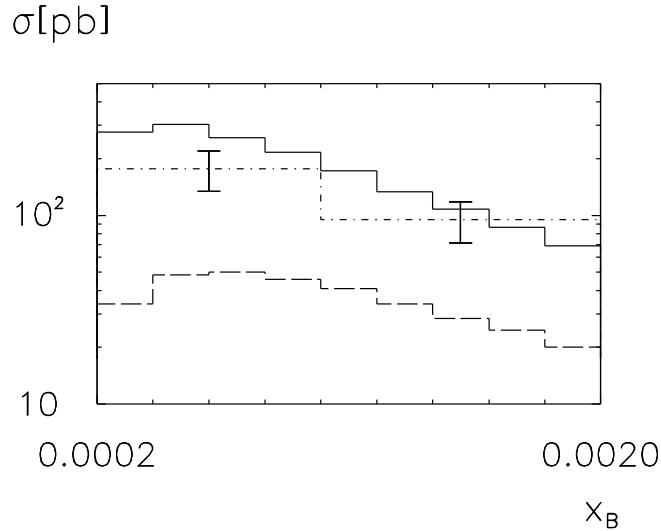,height=7.0cm}
\end{center}
\caption{Forward jet production: BFKL calculation (solid), approximate
three-parton matrix elements (dashed), and data of the H1 experiment
(dot-dashed).
\label{fig:bfklx}}
\end{figure}

The data of the H1 collaboration,\cite{H1forjet}
together with the results of a recent calculation
based on BFKL dynamics,\cite{BaDe96} are
shown in Fig.~\ref{fig:bfklx} as a function
of Bjorken $x$. The H1 selection criteria for
forward jets are $x_\jet>0.025$, $0.5<p^2_{T,\jet}/Q^2<4$,
$p_{T,\jet}>5$ GeV and $6^\circ<\theta_\jet<20^\circ$,
where $\theta_\jet$ is the angle between the forward jet
(defined by the cone algorithm) and the proton beam.
The agreement between the BFKL calculation and the
data is encouraging. The approximate matrix element
calculation shown, performed by omitting the BFKL ladder
in Fig.~\ref{fig:bfkl_jet}, falls well below the data.
It will be important to check that this remains true
when the full NLO matrix elements are included.
For this purpose the new NLO programs discussed
above, which allow the cone jet definition to be used,
will be most valuable.


\section*{Acknowledgments}
I am grateful to many colleagues, especially Yu.L.\ Dokshitzer,
G.\ Marchesini, A.H.\ Mueller and G.\ Salam, for helpful
discussions on the topics of this talk.
It is also a pleasure to thank the organizers of DIS96 for arranging
a most stimulating meeting in such a wonderful location.
This work was supported in part by the U.K.\ Particle Physics and Astronomy
Research Council and by the E.C.\ Programme ``Human Capital and Mobility'',
Network ``Physics at High Energy Colliders'', contract
CHRX-CT93-0357 (DG 12 COMA).

\end{document}